\documentclass[useAMS,usegraphicx,usenatbib]{mn2e}

\usepackage{ulem}
\usepackage{color}
\usepackage{graphics,graphicx}
\usepackage{times} 
\usepackage{amssymb}
\usepackage{amsmath}
\usepackage{lscape}
\usepackage{url}
\usepackage{multirow}
\usepackage{ctable}
\usepackage{sidecap}
\usepackage{rotating}
\usepackage{threeparttable}
\usepackage{float}
\usepackage{lscape}
\newif\ifAMStwofonts
\AMStwofontstrue
\title[Searching for Winds in ULXs: the Fe K band]{Searching for Massive Outflows in Holmberg\,IX X-1 and NGC\,1313 X-1: The Iron K Band}

\author[D.~J. Walton, J.~M. Miller, R.~C. Reis \& A.~C. Fabian]
{D.~J. Walton$^{1}$ \thanks{E-mail: dwalton@ast.cam.ac.uk},
J.~M. Miller$^{2}$,
R.~C. Reis$^{2}$ and
A.~C. Fabian$^{1}$ \\
\footnotesize
$^1$ Institute of Astronomy, Cambridge University, Madingley Road, Cambridge, CB3 0HA \\
$^2$ Department of Astronomy, University of Michigan, 500 Church Street, Ann Arbor, MI 48109, USA}
\date{}

\def\xmm{{\it XMM-Newton}}

\def\epicpn{{\it EPIC}{\rm-pn}}
\def\epicmos1{{\it EPIC}{\rm-MOS1~\/}}
\def\epicmos2{{\it EPIC}{\rm-MOS2 ~\/}}
\def\epicmos{{\it EPIC}{\rm-MOS}}

\def\kms{\hbox{$\rm\thinspace km~s^{-1}$}}

\def\H0{{\rm km s$^{-1}$ Mpc$^{-1}$}}

\def\kev{\hbox{$\rm\thinspace keV$}}

\def\atpcm{{\rm atom~cm$^{-2}$}}

\def\ergps{\hbox{erg~s$^{-1}$}}

\def\le{$L_{\rm E}$}

\def\msun{\hbox{$\rm\thinspace M_{\odot}$}}

\def\chisq{{$\chi^{2}$}}
\def\rchi{{$\chi^{2}_{\nu}$}}

\def\xspec{\hbox{\small XSPEC~\/}}
\def\xspecv{\hbox{\small XSPEC}\thinspace v12.7.0r}

\def\xmmselect{\hbox{\rm{\small XMMSELECT}}}
\def\ftool{\hbox{\rm{\small FTOOL}}}

\def\addspec{\hbox{\rm{\small ADDSPEC}}}

\def\grppha{\hbox{\rm{\small GRPPHA~\/}}}

\def\sas{\hbox{\rm{\small SAS~\/}}}
\def\epchain{\hbox{\rm{\small EPCHAIN~\/}}}
\def\emchain{\hbox{\rm{\small EMCHAIN~\/}}}

\def\rmfgen{\hbox{\rm{\small RMFGEN}}}
\def\arfgen{\hbox{\rm{\small ARFGEN}}}

\def\diskpn{\rm{\small DISKPN~\/}}

\def\grid25{\hbox{\rm{\small GRID25}}}

\def\tbabs{\rm{\small TBABS}}

\def\comptt{\rm{\small COMPTT}}

\def\ka{$K\alpha$\thinspace}

\def\zsun{\hbox{$\rm Z_{\odot}$}}

\def\eg{{\it e.g.~\/}}
\def\etc{{\it etc.}}
\def\ie{{\it i.e.~\/}}

\def\la{\mathrel{\hbox{\rlap{\hbox{\lower4pt\hbox{$\sim$}}}{\raise2pt\hbox{$<$}}}}}
\def\ga{\mathrel{\hbox{\rlap{\hbox{\lower4pt\hbox{$\sim$}}}{\raise2pt\hbox{$>$}}}}}

\def\d25{D$_{25}$}
\def\nh{{$N_{\rm H}$}}

\def\.25{0.25 keV\thinspace}
\def\lx{$L_{\rm X}$}

\def\mbh{\rm $M_{\rm BH}$}

\def\rg{$R_{G}$}

\def\mdot{$\dot{M}$}

\begin{document}

\pagerange{\pageref{firstpage}--\pageref{lastpage}}
\pubyear{2010}

\maketitle

\label{firstpage}

\begin{abstract} We have analysed all the good quality \xmm\ data publicly available for
the bright ULXs Holmberg\,IX X-1 and NGC\,1313 X-1, with the aim of searching for discrete
emission or absorption features in the Fe K band that could provide observational evidence
for the massive outflows predicted if these sources are accreting at substantially
super-Eddington rates. We do not find statistically compelling evidence for any atomic lines,
and the limits that are obtained have interesting consequences. Any features in the immediate
Fe K energy band (6--7\,\kev) must have equivalent widths weaker than $\sim$30\,eV for
Holmberg\,IX X-1, and weaker than $\sim$50\,eV for NGC\,1313 X-1 (at 99 per cent confidence).
In comparison to the sub-Eddington outflows observed in GRS\,1915+105, which imprint iron
absorption features with equivalent widths of $\sim$30\,eV, the limits obtained here appear
quite stringent, particularly when Holmberg\,IX X-1 and NGC\,1313 X-1 must be expelling at
least 5--10 times as much material if they host black holes of similar masses. The difficulty
in reconciling these observational limits with the presence of strong line-of-sight outflows
suggests that either these sources are not launching such outflows, or that they must be
directed away from our viewing angle.
\end{abstract}


\section{Introduction}

Ultraluminous X-ray Sources (ULXs) are extra-nuclear point sources in external galaxies
observed to be more luminous in X-rays than the Eddington luminosity \le\ for a stellar
mass ($\sim$10\msun) black hole (\lx\,$\gtrsim$\,$10^{39}$\,\ergps); on rare occasions
ULX luminosities have been observed to exceed $10^{41}$\,\ergps\ (\citealt{WaltonULXcat,
Sutton12}) and even to reach as high as $\sim10^{42}$\,\ergps\ (\citealt{Farrell09}).
This combination has led to extended debate over the nature of these sources. The
majority of early theories were based around either the presence of intermediate mass
black holes (IMBHs: $10^{2}$ $\lesssim$ \mbh\ $\lesssim$ $10^{5}$\,\msun;
\citealt{Colbert99}), super-Eddington accretion states for standard mass binary systems
(\eg \citealt{Poutanen07}, \citealt{Finke07}) or the inferred luminosities being
artificially high due to anisotropic emission (\citealt{King01}), although observations
of excited emission lines, typically He {\small II} $\lambda$4686, from either the
accretion disc of the ULX or its surrounding nebula appear to rule out highly beamed
emission via photon counting arguments (\eg \citealt{Pakull02}, \citealt{Kaaret04}, and
\citealt{Berghea10}).

More recently, proposed interpretations for bright ULXs (\lx\,$\sim$\,$10^{40}$
\ergps) tend to incorporate elements from each of these three ideas, invoking black
holes only slightly larger than those seen in Galactic binary systems (\mbh\
$\lesssim$ 100\,\msun) with mildly super-Eddington accretion rates (luminosities up
to a few times \le), such that the accretion disc is geometrically thicker than the
standard thin discs predicted for more moderate accretion rates (\citealt{Shakura73})
and forces some mild anisotropy (`slim' discs; see \eg \citealt{Abram80}). For recent
reviews on the observational status and the potential nature of ULXs see
\cite{Roberts07rev} and \cite{Feng11rev}.

\begin{table*}
  \caption{Basic details of the two bright ULXs considered in this work, Holmberg\,IX
X-1 and NGC\,1313 X-1, and the \xmm\ observations analysed. Distances are taken from
\citet{Paturel02} and \citet{TULLY} respectively.}
\begin{center}
\begin{tabular}{c c c c c c c c c} 
\hline
\hline
\\[-0.3cm]
Source & RA & DEC & $D$ & $z$ & OBSID & Obs. Date & Duration & Good Exposure \\
& (h:m:s) & (d:m:s) & (Mpc) & & & & (ks) & (\epicpn; ks) \\
\\[-0.3cm]
\hline
\hline
\\[-0.25cm]
Holmberg\,IX X-1 & 09:57:53.2 & +69:03:48.3 & 3.55 & 0.000153 & 0112521001 & 10/04/2002 & 11 & 7 \\
& & & & & 0112521101 & 16/04/2002 & 12 & 8 \\
& & & & & 0200980101 & 26/09/2004 & 119 & 88 \\
\\[-0.15cm]
NGC\,1313 X-1 & 03:18:20.0 & -66:29:11.0 & 3.7 & 0.001568 & 0106860101 & 17/10/2000 & 42 & 27\\
& & & & & 0150280601 & 08/01/2004 & 55 & 10 \\
& & & & & 0150181101 & 16/01/2004 & 9 & 6 \\
& & & & & 0205230201 & 01/05/2004 & 13 & 3 \\
& & & & & 0205230301 & 05/06/2004 & 12 & 9 \\
& & & & & 0205230601 & 07/02/2005 & 14 & 10 \\
& & & & & 0405090101 & 15/10/2006 & 123 & 99 \\
\\[-0.25cm]
\hline
\hline
\end{tabular}
\label{tab_obs}
\end{center}
\end{table*}

One of the observational results in recent X-ray studies of ULXs that has sparked
significant interest is that many of these sources display curvature in their X-ray
continuum above $\sim$3\,\kev\ (\citealt{Stobbart06, Gladstone09, Walton4517}).
Although some of these are lower luminosity sources that display (hot) thermal
disc-like spectra, in which high energy curvature would naturally be expected, even
the brighter ULXs with two apparently distinct emission components, potentially
analogous to the disc and Comptonised corona observed in Galactic black hole binaries
(BHBs), also appear to show curvature in their high energy components. The hard,
Comptonised emission rarely displays similar curvature in the standard sub-Eddington
accretion states of BHBs (for a recent review see \citealt{Remillard06rev}), and if
this is an intrinsic difference in the high energy continuum, it could imply that
there are fundamental differences between the accretion onto ULXs and their lower
luminosity BHB cousins. Given that the basic accretion geometry is not expected to
evolve with black hole mass, this might indirectly suggest we are not viewing
sub-Eddington accretion.

In this high-Eddington framework, \cite{Gladstone09} propose that the high energy
curvature could be due to Comptonisation in an optically thick corona, which
shrouds the inner accretion disc. The effect of the corona causes the observed
temperature of the disc (identified here with the soft component) to appear
artificially low as only the outer disc is observed directly. Low disc temperatures
have frequently been inferred for bright ULXs, which if otherwise associated with
the inner disc would imply the presence of IMBHs (see \eg \citealt{MilFab04}). This
interpretation has been slightly modified by \cite{Middleton11a}, who propose that
the cool, quasi-thermal emission previously associated with the disc is instead
thermalised emission from a photosphere associated with the base of an outflowing
wind, as strong outflows are ubiquitously predicted by models for high and
particularly super-Eddington accretion (see \citealt{Ohsuga11}, and references
therein).

Galactic BHBs frequently display evidence for mass outflows, which at moderately
high accretion rates (during the thermal dominated states) take the form of
equatorial disc winds with outflow velocities $v_{\rm out} \lesssim 1000$\,\kms.
Prominent examples are the BHBs GRO\,J1655-40 (\citealt{Miller06a, DiazTrigo07})
and GRS\,1915+105 (\citealt{Kotani00, Lee02, Ueda09, Neilsen09}). When these
outflows impinge upon our line of sight to the central source, their observational
consequence is to imprint absorption features onto the intrinsic X-ray continuum,
typically in the form of narrow absorption lines from highly ionised species, with
the most prominent features arising from the \ka\ transitions of highly ionised
iron (Fe {\small XXV} and/or {\small XXVI}). In the case of GRS\,1915+105, perhaps
the best studied BHB outflow, \cite{Neilsen09} show that the strength of the
absorption features generally increases as the luminosity of the source, and hence
its Eddington ratio increases. The mass of this source is relatively well known
(\mbh\ = $14\pm4$\,\msun; \citealt{Greiner01}), and it is observed to radiate up to,
and possibly slightly in excess of its Eddington limit (\citealt{Vierdayanti10grs}).
While the geometry of the BHB disc winds observed at moderate accretion rates are
largely equatorial, the scale height of these winds is generally expected to
increase with Eddington ratio (see \eg \citealt{Abramowicz05}, \citealt{King09}),
so at higher accretion rates the outflows should probably cover a larger solid angle.

If instead, super-Eddington outflows are present in ULXs but, despite their potential
large scale heights, do not occur along our line of sight, rather than viewing
absorption features we should instead observe reprocessed emission from the material
in the outflow. Indeed, reprocessed emission lines consistent with neutral (or
moderately ionised) iron are seen ubiquitously in the spectra of Galactic high mass
X-ray binaries (HMXBs; see {\citealt{Torrejon10}), which are expected to accrete
largely from the stellar winds launched by their massive companions. Furthermore,
the majority of ULXs themselves are widely expected to be HMXB analogues, owing to
the their apparent correlation with recent star formation (\citealt{Swartz09}).
Recent population studies of ULXs appear to support this picture, as their luminosity
function appears consistent with being a smooth extrapolation of the HMXB luminosity
function (at least for sources in spiral/irregular type galaxies; \citealt{Grimm03},
\citealt{WaltonULXcat}, \citealt{Swartz11}). 

Here, we present an analysis of the Fe K band for two bright ($L_{\rm X} \sim
10^{40}$\,\ergps) ULXs with hard X-ray spectra, Holmberg\,IX X-1 (also known as M\,81
X-9) and NGC\,1313 X-1, in which we search for evidence of atomic iron features that
could be associated with outflowing material. These sources have been selected as
they represent the datasets with the highest quality data at high energies for such
bright, isolated ULXs. The paper is structured as follows: section \ref{sec_obs}
describes our data reduction prodecure, and section \ref{sec_analysis} describes the
analysis performed. We discuss our results in section \ref{sec_discussion} and
summarise our conclusions in section \ref{sec_conc}.

\section{Observations and Data Reduction}
\label{sec_obs}

\begin{table*}
  \caption{Key parameters obtained for Holmberg\,IX X-1 and NGC\,1313 X-1 with the
phenomenological ultraluminous state continuum model (see text).}
\begin{center}
\begin{tabular}{c c c c c c c c c c} 
\hline
\hline
\\[-0.3cm]
Source & Obs. & \nh & $kT_{\rm in}$ & $kT_{\rm e}$ & $\tau$ & \rchi\ ($\chi^{2}$/d.o.f.) \\
& & ($10^{21}$ cm$^{-2}$) & (keV) & (keV) &  \\
\\[-0.3cm]
\hline
\hline
\\[-0.25cm]
Holmberg\,IX\,X-1 & Combined & $1.5\pm0.1$ & $0.23\pm0.01$ & $2.4\pm0.1$ & $8.7\pm0.3$ & 1.08 (1998/1851) \\
\\[-0.35cm]
& 0200980101 & $1.5\pm0.1$ & $0.23\pm0.01$ & $2.3\pm0.1$ & $9.2\pm0.4$ & 1.02 (1804/1762) \\
\\[-0.15cm]
NGC\,1313\,X-1 & Combined & $2.5\pm0.1$ & $0.22\pm0.01$ & $2.3^{+0.2}_{-0.1}$ & $7.6\pm0.4$ & 1.16 (1867/1608) \\
\\[-0.35cm]
& 0405090101 & $2.6^{+0.2}_{-0.1}$ & $0.22\pm0.01$ & $2.2^{+0.1}_{-0.2}$ & $8.5\pm0.5$ & 1.08 (1415/1305) \\
\\[-0.25cm]
\hline
\hline
\end{tabular}
\label{tab_comptt}
\end{center}
\end{table*}

In this work we make use of all the publically available \xmm\ observations with good
\epicpn\ (\citealt{XMM_PN}) and \epicmos\ (\citealt{XMM_MOS} spectra of the two bright
ULXs considered, Holmberg\,IX X-1 and NGC\,1313 X-1. In the latter case, there are a
number of observations pointed at NGC\,1313 X-2 in which X-1 fell on a chip gap in one
or more of the \xmm\ detectors which were not included in our analysis. In addition,
we only consider observations in which the respective sources are detected reliably
above background up to 10\,\kev. Data reduction was carried out with the Science
Analysis System (\sas v11.0.0) largely according to the standard prescription provided
in the online guide\footnote{http://xmm.esac.esa.int/}.

Observation data files were processed using \epchain and \emchain to produce calibrated
event lists for the \epicpn\ and \epicmos\ detectors respectively. Spectra were produced
separately for each observation by selecting only single and double events (single to
quadruple events) for \epicpn\ (\epicmos) using \xmmselect, and periods of high background
were treated according to the method outlined by \cite{Picon04} with the signal-to-noise
ratio maximised for the full energy band considered (0.3--10.0\,\kev). Circular
source regions were used, and although the exact size varied from observation to
observation depending on the location of chip gaps, \etc\ these were typically of radius
$\sim$30--40''. Larger circular regions typically of radius $\sim$50'' and $\sim$75'' free
of other sources and on the same CCD as the ULX were selected in order to sample the
background for the \epicpn\ and \epicmos\ detectors respectively. Redistribution matrices
and auxiliary response files were generated with \rmfgen\ and \arfgen. After performing
the data reduction separately for each of the observations, the spectra were then combined
using the \ftool\footnote{http://heasarc.nasa.gov/ftools/ftools\_menu.html} \addspec\ in
order to maximise the data quality in the Fe K band. Finally, \epicpn\ and \epicmos\
spectra were re-binned using \grppha to have a minimum of 25 counts in each energy bin,
so that the probability distribution of counts within each bin can be considered Gaussian,
and hence the use of the $\chi^2$ statistic is appropriate when performing spectral fits.

The observations that met our selection criteria and are hence included in our analysis
are listed in Table \ref{tab_obs}; the resulting good exposure times obtained in the stacked
spectra are 103 and 238\,ks with the \epicpn\ and combined \epicmos\ detectors respectively
for Holmberg\,IX X-1, and 162 and 381\,ks for NGC\,1313 X-1, although these are largely
dominated by the single full orbit observations available for each source (contributing
88/200\,ks to the \epicpn/\epicmos\ spectra of Holmberg\,IX X-1, and 90/215\,ks to the
spectra of NGC\,1313 X-1 respectively).

\section{Spectral Analysis}
\label{sec_analysis}

Throughout this work, spectral modelling is
performed with \xspecv\ (\citealt{XSPEC}), and all quoted uncertainties are the 90 per cent
confidence limits for a single parameter of interest, unless stated otherwise. In the
following, Holmberg\,IX X-1 and NGC\,1313 X-1 are modelled over the full 0.3--10.0\,\kev\
energy range. The \epicpn\ and \epicmos\ spectra are modelled simultaneously with all
parameters tied between the two spectra, although we do include a multiplicative constant
in the \epicmos\ spectrum in order to account for possible cross-calibration uncertainties
between the detectors. This value is always found to be within $\sim$10 per cent of unity.

\begin{figure*}
\begin{center}
\hspace*{-0.4cm}
\rotatebox{0}{
{\includegraphics[width=230pt]{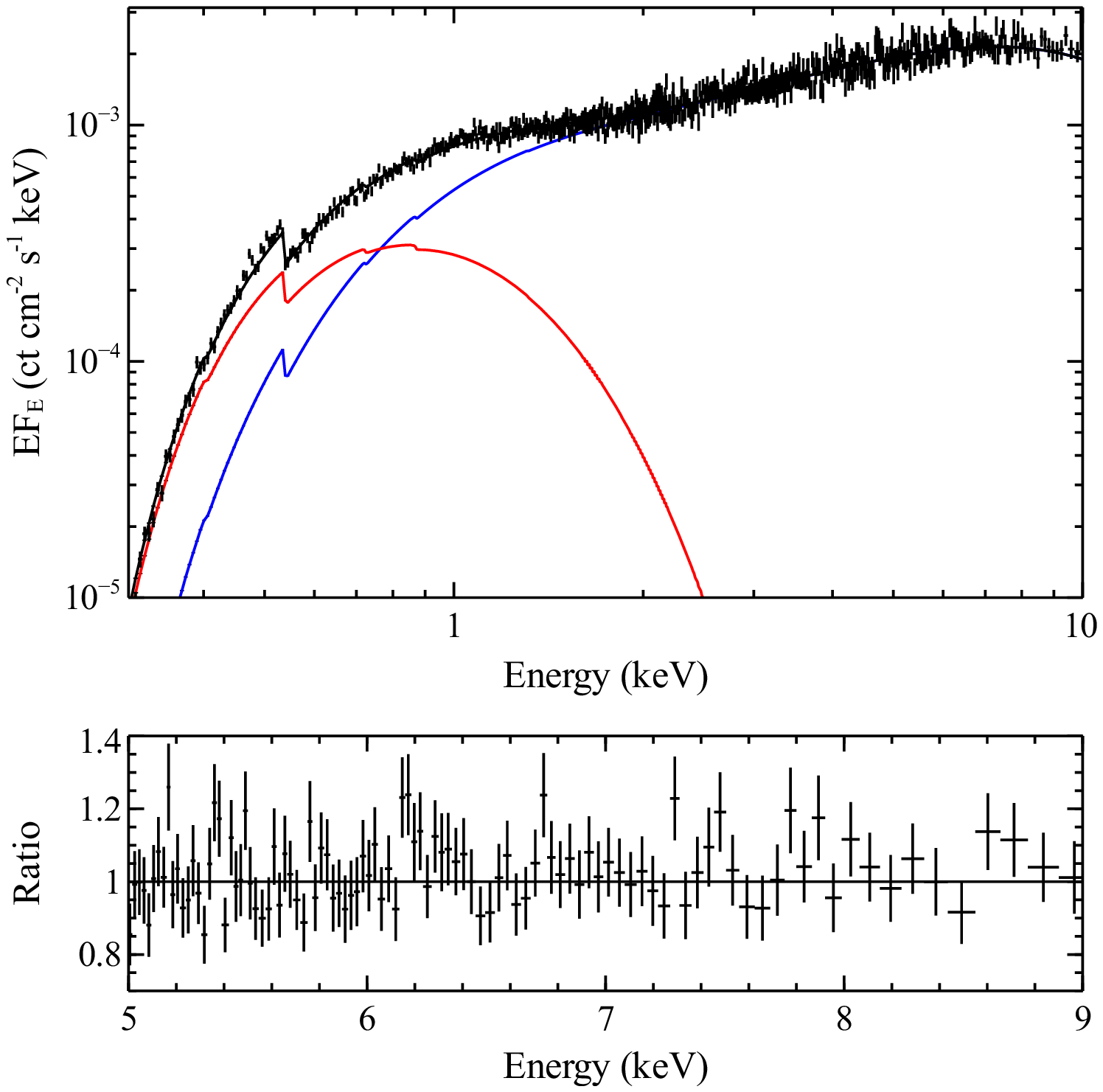}}
}
\hspace{0.75cm}
\rotatebox{0}{
{\includegraphics[width=230pt]{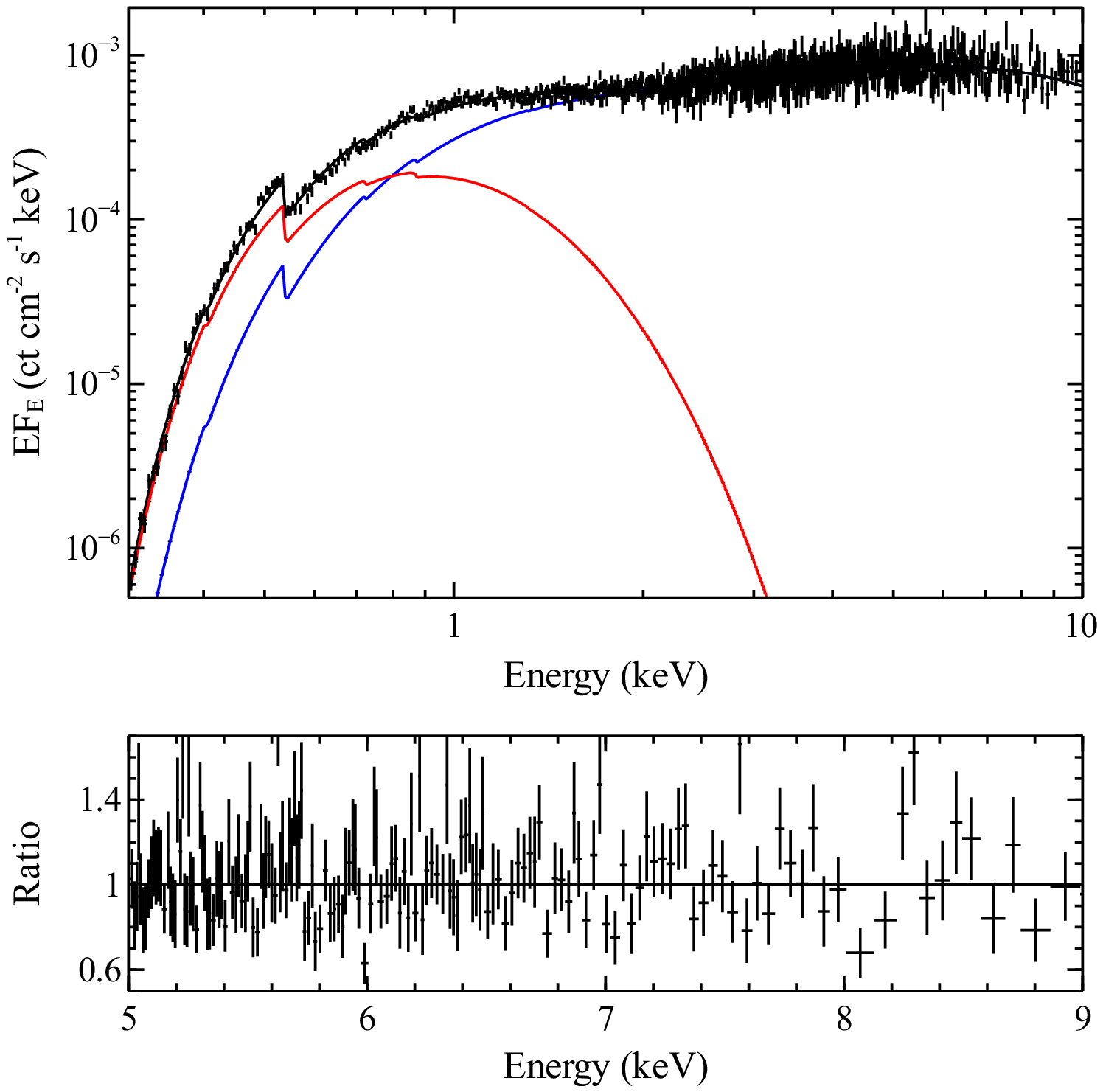}}
}
\hspace{0.5cm}
\end{center}
\caption{Continuum model fits to the stacked spectra of Holmberg\,IX X-1 (\textit{left
panel}) and NGC\,1313 X-1 (\textit{right panel}). The models consist of a cool disc
component (red) and cool, optically thick Comptonisation (blue), in accordance with the
ultraluminous state concept for super-Eddington accretion proposed by \citet{Gladstone09}.
The lower panels show the data/model ratio for the models constructed over the energy
range of interest for this work, 5--9\,\kev; these models clearly provide a good
representation of the high energy continuum.}
\label{fig_fits}
\end{figure*}

\subsection{Continuum Model}
\label{sec_cont}

We begin by modelling the stacked spectra of these sources with a simple phenomenological
interpretation of the ultraluminous state concept, including a cool disc (\diskpn;
\citealt{diskpn}) to model the soft component and cool, optically thick thermal
Comptonisation (\comptt; \citealt{comptt}) to model the curvature of the hard component.
The inner radius of the disc is fixed at 6\,\rg\ (= GM/c$^{2}$) and the seed photon
temperature for \comptt\ is linked to the inner temperature of the disc. Both components
are modified by neutral absorption with solar abundances (\tbabs; \citealt{tbabs}).
The overall quality of fit obtained over the full 0.3--10.0\,\kev\ band is generally
fairly good, with \rchi\ = 1998/1851 and 1867/1608 for the stacked spectra of
Holmberg\,IX X-1 and NGC\,1313 X-1 respectively, but if we focus on the quality of fit
obtained with these models (without refitting) at higher energies (2--10\,\kev) we find
it to be much improved, with \rchi\ = 1429/1394 and 1218/1151. We therefore conclude that
the continuum is well modelled over the energy band of interest for this study
($\sim$5--9\,\kev) for both sources, see Fig. \ref{fig_fits}. The results obtained for
the key continuum parameters are quoted in Table \ref{tab_comptt}. For comparison, we
also repeat our analysis for the long observations of each source separately; we obtain
good agreement with the results initially presented in \cite{Gladstone09}.

To confirm whether it is reasonable to combine the individual observations into a single
spectrum, we also apply this continuum model to each of the observations individually.
For each source we model all the individual observations considered simultaneously,
requiring that the neutral absorption remains the same for each observation, but allowing
the other key parameters to vary (see Appendix \ref{appendix} for full details). The key
parameters obtained for each of the observations are all quite similar, therefore we
conclude it is reasonable to combine the observations included in this work.


\subsection{Fe K Band}

Having successfully reproduced the shape of the continuum for these two sources, we now
proceed to search for evidence of narrow iron emission or absorption features which might
indicate a large, super-Eddington outflow. We are primarily interested in iron in this
work owing to its high cosmic abundance and fluorescent yield, and also owing to the fact
that it is difficult to fully ionise. Furthermore, the \xmm\ response function varies
slowly and smoothly over the iron K$\alpha$ bandpass. To search for these features, we
add a narrow ($\sigma = 10$\,eV) Gaussian feature with a normalisation that can either be
positive to model potential emission features, or negative to model potential absorption
features. The restframe energy of this component (taking into account the redshift
adopted for each source, as quoted in Table \ref{tab_obs}) is systematically varied
across the energy range 5--9\,\kev\ in 100 steps, in order to incorporate the key iron
transitions and also (generously) allow for any potential blue shifts, as would be
expected for absorption features from a strong outflow. This approach is broadly similar
to that adopted in the recent works by \cite{Tombesi10b,Tombesi10a} during their initial
search for potential emission and absorption features in the high energy spectra of
active galaxies. At each energy increment we record the $\Delta\chi^{2}$ improvement
provided by the inclusion of the Gaussian feature, as well as the equivalent width ($EW$)
of any potential feature at that energy, and its 90 and 99 per cent confidence
limits\footnote{Note that these are confidence limits on line strength, which are not
strictly the same as the upper limits on the strength of lines that should be detected at
a given detection confidence level (see \citealt{Kashyap10}), but are simpler to calculate
in this instance, and should most likely provide more conservative estimates for the
maximum strength of any lines intrinsically present.}. These are obtained with the {\small
EQWIDTH} command in \xspec\ by generating 10000 parameter sets based on the best fit model
and the covariance matrix, which contains information on the parameter uncertainties
obtained with this fit, and extracting the 90 and 99 per cent uncertainty limits from the
distribution of equivalent width values obtained from each of these parameter sets.

The results obtained for Holmberg\,IX X-1 and NGC\,1313 X-1 are shown in Fig. \ref{fig_hoix}
and Fig. \ref{fig_ngc1313} respectively; in each case the left panel displays the results
obtained with the stacked spectrum, and the right panel displays the results obtained when
considering only the longest observation of the source. The top panels display the
$\Delta\chi^{2}$ improvement as a function of line energy, multiplied by the sign of the
normalisation to differentiate between emission and absorption, and the lower panels display
the limits on the equivalent widths obtained. For clarity, we highlight the energies of the
\ka transitions of neutral iron, iron {\small XXV} and iron {\small XXVI} with dotted
vertical lines. We also show $EW = \pm$30\,eV with dashed horizontal lines, which roughly
represent the strongest absorption features observed in GRS\,1915+105 by \cite{Neilsen09},
although as noted by those authors the $EW$ values obtained in that work are actually likely
to be lower limits owing to their continuum fitting technique, and in GRO\,J1655-40 by
\cite{Miller06a}.

First of all, we wish to stress that we have not found any statistically significant
detections of any features, either in absorption or in emission. Although a few energies are
picked out by our analysis as offering the most improvement, this is never greater than
$\Delta\chi^{2} \sim 7$ over the energy range considered. Therefore our discussion
is restricted to the $EW$ limits that can be placed instead, although these are interesting
to consider in their own right. We find that the data are of sufficient quality to exclude
very strong ($EW \gtrsim 50$\,eV) iron features in either absorption or emission across a
large portion of energy range immediately around the Fe K transitions, although the
constraints naturally become less restrictive at very high energies ($\gtrsim$7.5--8.0\,\kev)
as the data quality declines. The high energy data quality is better for Holmberg\,IX X-1,
so the limits that can be placed are strongest in this case, where any persistent atomic
features in the spectrum below $\sim$7.5\,\kev\ must have an equivalent width less than
$\sim$30\,eV. In the case of NGC\,1313 X-1, a moderately strong ($EW \lesssim 50$\,eV)
emission feature consistent with neutral iron and absorption feature consistent with iron
{\small XXVI} are both still permissable. In the following section we discuss the
implications of these limits on the potential presence of large outflows and the accretion
geometry for these ULXs.

\begin{figure*}
\begin{center}
\rotatebox{0}{
{\includegraphics[width=235pt]{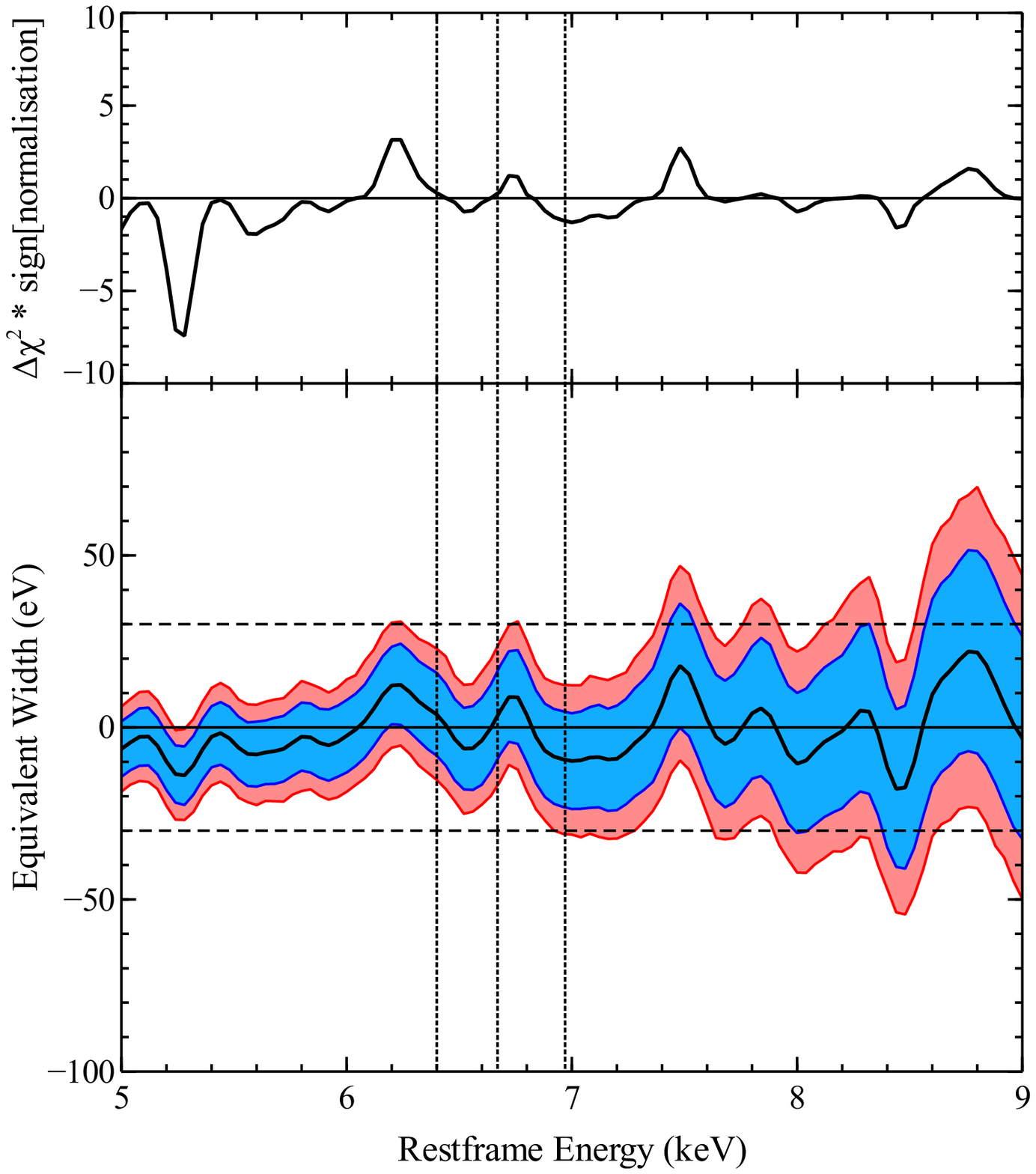}}
}
\hspace{0.25cm}
\rotatebox{0}{
{\includegraphics[width=235pt]{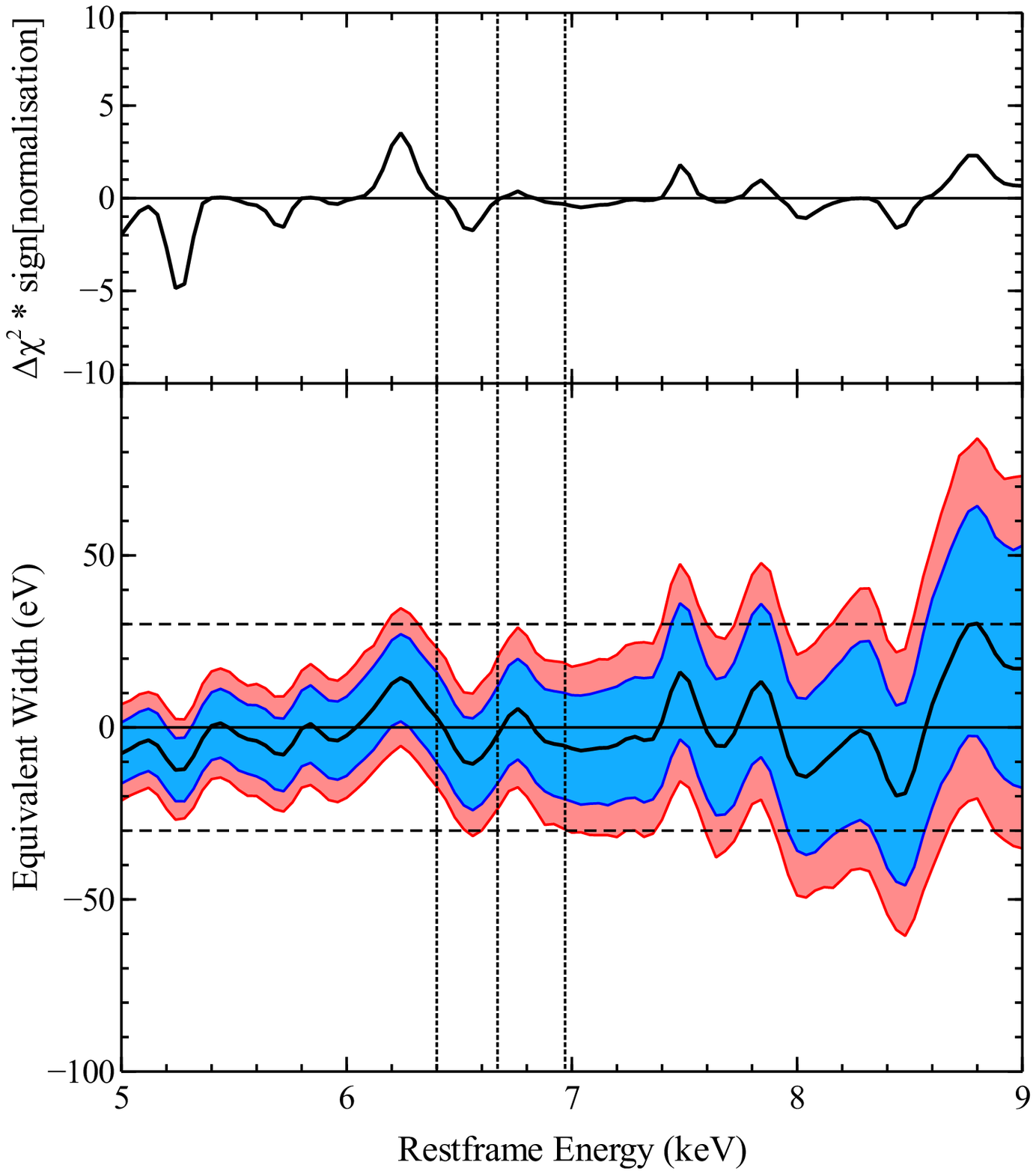}}
}
\end{center}
\caption{\textit{Top panels:} the $\Delta$\chisq\ improvement provided by including a narrow
($\sigma = 10$\,eV) Gaussian line, as a function of line energy, for Holmberg\,IX X-1. The
normalisation of the line is allowed to be either positive or negative to allow for both
emission and absorption features, and the $\Delta$\chisq\ improvement is multiplied by the
sign of the best fitting line normalisation to indicate whether the improvement is provided
by an emission or an absorption feature. We find no statistically compelling evidence for any
narrow iron features at high energies. \textit{Bottom panels:} 90 (\textit{blue}) and 99
(\textit{red}) per cent confidence contours for the equivalent width of the included narrow
line as a function of line energy, indicating the line strengths any persistent narrow
features intrinsically present could have and still remain undetected with the current data.
The rest frame transitions of neutral, helium-like and hydrogen-like iron (6.4, 6.67 and 6.97
keV) are shown with vertical dotted lines, and for comparison we also plot dashed horizontal
lines representing $EW = \pm$\,30\,eV. This analysis is repeated for both the stacked spectrum
(\textit{left panels}) and the longest single \xmm\ observation (\textit{right panels}).}
\label{fig_hoix}
\end{figure*}

\begin{figure*}
\begin{center}
\rotatebox{0}{
{\includegraphics[width=235pt]{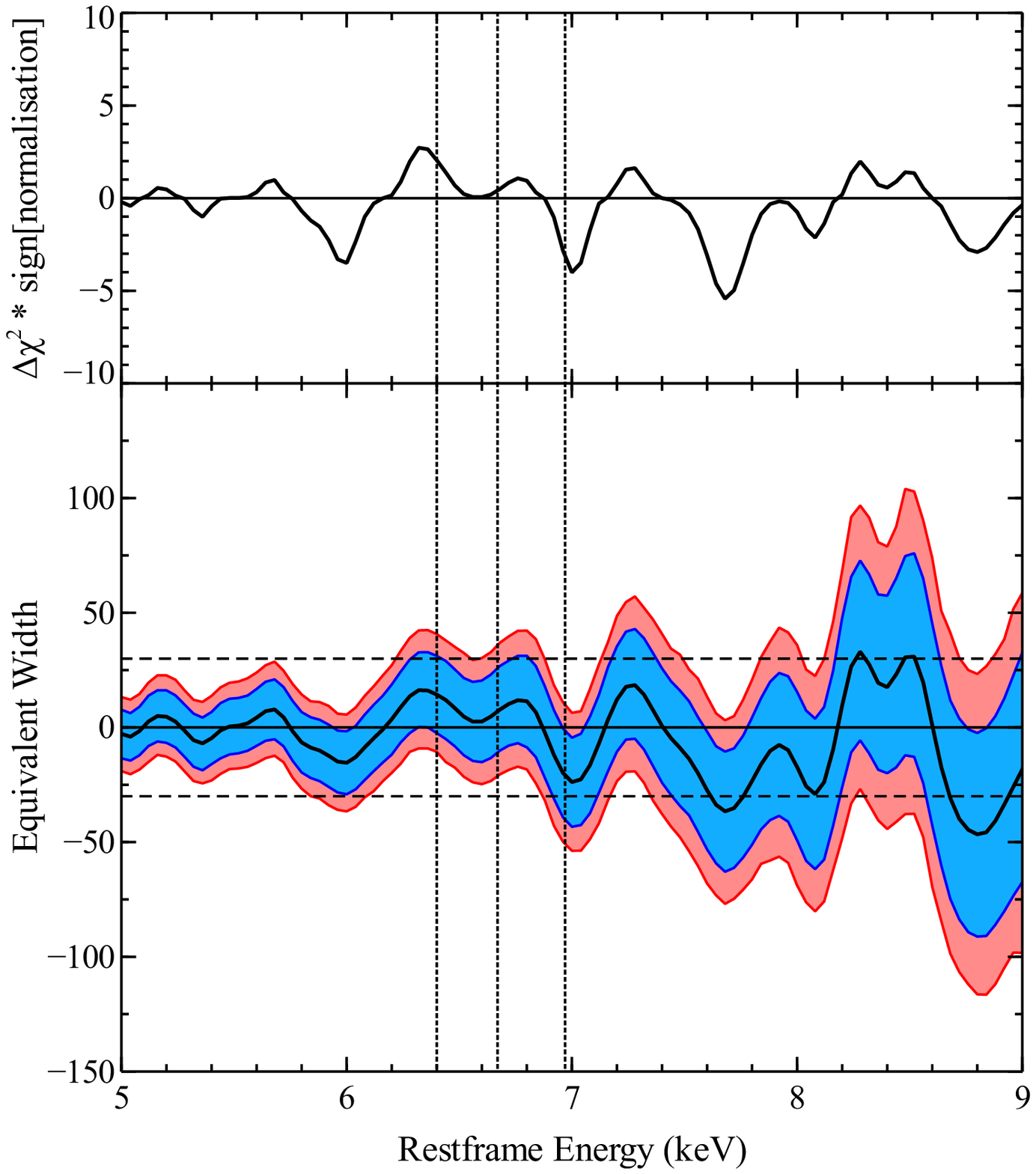}}
}
\hspace{0.25cm}
\rotatebox{0}{
{\includegraphics[width=235pt]{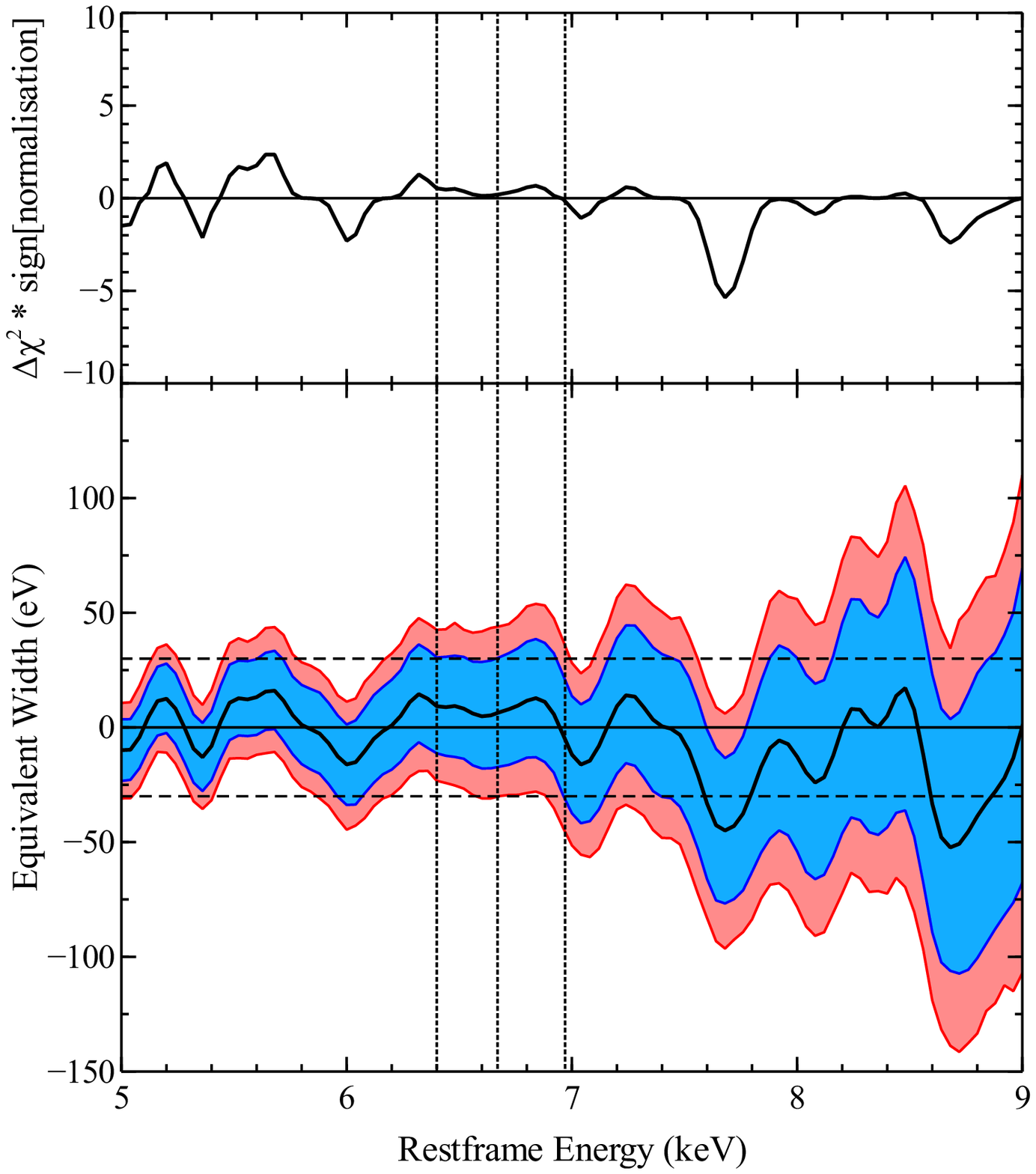}}
}
\end{center}
\caption{As in Fig. \ref{fig_hoix}, but here for NGC\,1313 X-1.}
\label{fig_ngc1313}
\end{figure*}

\section{Discussion}
\label{sec_discussion}

We have analysed the available \xmm\ observations for two bright ULXs with hard
X-ray spectra, Holmberg\,IX X-1 and NGC\,1313 X-1, with the purpose of searching
for direct observational evidence of massive super-Eddington outflows via atomic
features in the iron K band. In order to maximise the high energy data quality we
stacked the available data with simultaneous \epicpn\ and \epicmos\ coverage, and
proceeded to search for narrow features in either absorption or emission. We do
not find any statistically compelling evidence for any narrow features, so
instead we examine the limits that can be placed on any features intrinsically
present in the spectrum that remain undetected. For comparison we also analyse
the full orbit observations available for these two sources separately, which
make up $\sim$85 and 55 per cent of the total good exposure time analysed for
Holmberg\,IX X-1 and NGC\,1313 X-1 respectively. In the latter case the additional
observations contribute more significantly to the stacked spectrum, and a more
noticable improvement is obtained with their inclusion. However, the results are
similar in both cases, and we focus our discussion on the limits obtained with the
stacked spectra.

\subsection{Absorption Lines}

First we consider the limits that can be placed on persistent absorption features,
given the strong outflows expected from super-Eddington accretion. For Holmberg\,IX
X-1 any absorption features in the immediate Fe K band (6--7\,\kev) must have an
equivalent width less than $\sim$30\,eV, based on the more conservative 99 per cent
limits obtained. The constraints are not so strong for NGC\,1313 X-1, where an
absorption feature consistent with Fe {\small XXVI} could potentially still have an
equivalent width of up to $\sim$50\,eV. The implication is that any line-of-sight
outflows present cannot have any greater impact on the spectrum of Holmberg\,IX X-1
than the strongest sub-Eddington Galactic BHB outflows, observed in GRS\,1915+105
and GRO\,J1655--40, and can only have at most a moderately stronger impact for
NGC\,1313 X-1. Furthermore, we also draw attention to the outflow in the Galactic
BHB candidate IGR\,J17091--3624 (\citealt{King12}). This source displays strong
similarities to GRS\,1915+105 (\citealt{Altamirano11}), and though its distance and
mass are not currently known, through this comparison it has tentatively been
suggested that it could also be emitting at close to, or even in excess of its
Eddington limit. Although the statistical significances of the iron features in this
source are not particularly high, if real their reported equivalent widths are
$\sim$40 and $\sim$90\,eV for Fe {\small XXV} and {\small XXVI} when using a cool,
optically thick Comptonisation model for the continuum (similar to that
observed in GRS\,1915+105, and to the ultraluminous state interpretation). Any
undetected features in the immedite Fe K band must be much weaker than these.

However, assuming black hole masses of 10\,\msun\ for Holmberg\,IX X-1 and NGC\,1313
X-1, these sources are observed to persistently radiate at $\sim$5--10 times their
Eddington limits. In this scenario, we would expect them to be ejecting at least
5--10 times the amount of material as a similar source accreting at the Eddington
limit (\ie GRS\,1915+105). \cite{Ponti12} find that, even under the assumption of a
constant wind opening angle, the ratio of the mass outflow rate to the innermost
accretion rate, \mdot$_{\rm out}$/\mdot$_{\rm acc}$, increases with increasing
Eddington ratio, so increasing the luminosity by a factor of 10 increases the amount
of material expelled by an even larger factor (assuming this increase in luminosity
relates to an equivalent increase in \mdot$_{\rm acc}$). Given that $EW$ scales
roughly linearly with absorbing column density, if the ionisation state and
abundances of the winds remained constant we might therefore have expected to see
iron absorption features with $EW \gtrsim 300$\,eV if these sources are at $\sim$10
times the Eddington limit and their associated outflows are along our line-of-sight,
assuming solar abundances. For the $Z \sim 0.5$\,\zsun\ abundances of the host
galaxies in these cases (\citealt{Makarova02,Hadfield07,Pintore12}), the expectation
would still be $EW \gtrsim 150$\,eV.

If these outflows are truly absent, this would argue very strongly against a highly
super-Eddington interpretation for these ULXs. Both sources display hard X-ray spectra
with evidence for two separate continuum emission components, similar to some of the
complex `intermediate' sub-Eddington states, with very cool disc temperatures
($\sim$0.2\,\kev) obtained when associating the soft component with the accretion disc.
However, both sources also display evidence for the high energy curvature in their hard
components that might distinguish ULXs from the typical sub-Eddington states seen in
BHBs, and is potentialy the key signature of the proposed `ultraluminous' accretion
state (\citealt{Gladstone09}). In fact, GRS\,1915+105 may show some similar spectral
characteristics, including the high energy curvature, when radiating close to its
Eddington limit (\citealt{Vierdayanti10grs}), although the presence of a reflection
component in this source complicates the issue. If the predicted outflows for strong
super-Eddington accretion really are absent, this curvature might still indicate
another high Eddington-ratio accretion state, but one that is still roughly Eddington
limited rather than strongly super-Eddington, and observed when $L/L_{\rm E} \sim 1$.
This would suggest the black hole masses of Holmberg\,IX X-1 and NGC\,1313 X-1 are
$\sim$50--100\,\msun.

There are, however, some important caveats to these limits that must be considered,
especially given the ionised nebulae that many bright ULXs are observed to reside
within. In some cases, including Holmberg\,IX X-1, the outer regions of these
nebulae appear to be more consistent with being ionised by shocks instead of X-ray
photoionisation, and these regions are observed to be expanding (\citealt{Pakull02,
Abolmasov07, Cseh12}). Where present, these shock-ionised regions have generally
been interpreted as potential evidence for the large scale impact on the ISM of
strong winds launched from the central ULX (although, of course, the mere presence
of an outflow from the accretion disc does not by itself imply super-Eddington
accretion). Historically, interpretations in which they are instead the expanding
remnants of the explosive event in which the ULX was formed have also been proposed
(see \eg \citealt{Roberts03}), but given the typical ages of the surrounding stellar
populations (see \eg \citealt{Grise11}), the nebulae appear more consistent with
the wind inflated scenario. Therefore, we now consider whether, despite the
absorption limits obtained from our X-ray analysis, we could still be viewing these
bright ULXs through strong, super-Eddington outflows.

First, we note briefly that when considering the stacked spectra we are strictly
obtaining limits on persistent features, and that the outflows observed in BHBs do vary
with time, broadly correlating with accretion state (see \eg \citealt{Miller06a},
\citealt{JMiller08}, \citealt{Neilsen09}, \citealt{Ponti12}). However, although the
sources do display some mild long term flux variability between the observations
considered here, it is certainly not clear that these variations are related to
accretion state changes, particularly given that the individual spectra obtained from
the observations included in this work all appear intrinsically similar (see appendix
\ref{appendix}). Therefore, a search for persistent features seems reasonable.
Furthermore, similar (if mildly weaker) constraints are obtained with the single
full-orbit observations of each source, in which virtually no short term variability
is observed (\citealt{Heil09}).

Although the limits obtained across the immediate Fe K band (6--7\,\kev) are fairly
stringent, they naturally become less restrictive at higher energies. Stronger
absorption features could still be present in the spectrum were they to arise
in winds with very high outflow velocities. This is particularly true for NGC\,1313
X-1 where persistent narrow absorption features with equivalent widths up to
$\sim$75\,eV (or even $\sim$100\,eV at truly extreme velocities) could still currently
be present. In contrast,  even high velocity features are constrained to have $EW
\lesssim 50$\,eV for Holmberg\,IX X-1. In either case, outflow velocities of $v_{\rm
out} \gtrsim$ 20,000\,\kms\ (0.08$c$) would be required in order to invoke absorption
features significantly stronger than currently possible in the immediate Fe K band.
To date, the fastest potential outflow observed in any BHB has a velocity of $v_{\rm
out} \sim$ 10,000\,\kms\ (IGR\,J17091--3624; \citealt{King12}), although much more
typical velocities are $v_{\rm out} \lesssim 1000$\,\kms. However, there is growing
evidence for winds with $v_{\rm out} \gtrsim 0.1c$ in active galaxies
(\citealt{Pounds09, Reeves09, Tombesi10b}). Indeed, the outflows required in order
to inflate the shocked regions of the observed nebulae may need to have mildly
relativistic outflow velocities (\citealt{Pakull02, Abolmasov07}), similar to those
seen in the Galactic system SS\,433 (see \citealt{Fabrika04} and references therein).
Outflows with extreme velocities therefore remain an intriguing possibility for ULXs,
particularly if IGR\,J17091--3624 turns out to be radiating at super-Eddington
luminosities. 

However, the most obvious possibility, assuming that these sources do launch strong,
super-Eddington outflows along our line of sight, is that in contrast to Galactic
BHBs, the ionisation state of the material is such that it is not readily observable
via iron absorption. The ionisation state of a photoionised plasma is typically
quantified as $\xi = L/nR^{2}$, where $L$ is the ionising luminosity from the source
at a distance $R$ and $n$ is the density of the plasma. The outflows in GRS\,1915+105
typically display ionisation parameters of $\log\xi \sim 4$. We know how $L$ changes
between GRS\,1915+105 and these bright ULXs, and therefore how the ionisation should
change for a constant outflow density and distance, but the combination of $nR^{2}$
is not easy to constrain and will determine how the ionisation truly evolves.

Observationally, a lower ionisation outflow does not seem particularly likely in these
cases as a massive (high column) outflow with relatively low ionisation should have
very obvious effects elsewhere in the spectrum, around the oxygen transitions
($\sim$0.5--1.0 \kev). Indeed, for brief illustration, we insert a moderately ionised
absorber using the {\small XSTAR} photoionisation code into the model used for
Holmberg\,IX X-1, fixing the ionisation at $\log\xi = 2$ and requiring that the
redshift of the absorber is either consistent or blueshifted with respect to the
redshift of the host galaxy. Applying this absorber to the full continuum model, we
find that the column density of any moderately ionised material must be \nh\ $\lesssim
6 \times 10^{20}$\,\atpcm\ (note that the model used has a lower limit of
$10^{20}$\,\atpcm\ to the allowed column densities), compared to the columns of $\sim$
10$^{23}$\,\atpcm\ seen in the sub-Eddington outflows of GRS\,1915+105 and GRO\,J1655-40.

It is therefore probably more likely, based on these observational constraints, that
the ionisation state of any line-of-sight wind is such that the bulk of the iron is
close to being fully ionised, and hence is prevented from imprinting any strong
features on the spectrum. Indeed, continuing their detailed study of the wind in
GRS\,1915+105, \cite{Neilsen12} present tentative evidence that the wind in this
source may start becoming overionised at the highest luminosities. However, we note
that at the ionisation states typically displayed by the wind in GRS\,1915+105,
increasing $\xi$ by a factor of $\sim$10, in line with the differences in luminosity
between this source and the ULXs considered here, only decreases the $EW$ of the iron
features that would be observed by a factor of $\sim$2 (for a constant column density),
so even accounting for the sub-solar metallicity of the host galaxy, this scenario
would require that the ionisation increase by a substantially larger factor than could
be accounted for simply by the difference in luminosity.

Such a large increase in ionisation could potentially occur through either a simultaneous
decrease in the distance between the ionising source and the outflow material, \ie the
launch radius of the wind decreasing, or through a simultaneous decrease in outflow
density. Given that we expect more material to be expelled at higher Eddington ratios,
we consider the latter evolution to be unlikely. However, the former would be contrary
to the larger radius invoked in the toy model for highly super-Eddington outflows
proposed by \cite{Middleton11b}, and the more general theoretical prediction that the
radius from which outflows are launched by radiation pressure increases with increasing
\mdot$_{\rm out} / $ \mdot$_{\rm acc}$ (\citealt{Shakura73}). Therefore, such a large
increase in ionisation appears difficult to justify on physical grounds. Furthermore, we
note that if IGR\,J17091--3624 does turn out to be a strongly super-Eddington source,
this overionised scenario would be ruled out on observational grounds.

Individually or in combination, all of these effects can increase the difficulty of
detecting strong line-of-sight outflows via iron absorption, even if such outflows are
present. Nonetheless, the limits that can currently be placed on any persistent iron
absorption features in the spectra of Holmberg\,IX X-1 and NGC\,1313 X-1, when
considered in the context of the behaviour of sub-Eddington BHBs, are clearly already
beginning to tell us something about their accretion/outflow processes. Any
interpretation invoking highly super-Eddington accretion with outflows along our line
of sight for these sources must be able to explain the surprising lack of strong iron
features in comparison to sub-Eddington BHBs given the expected increase in outflow
rates. However, given the apparent difficulty in explaining the increase in outflow
ionisation required for the geometry in which the outflows cross our line of sight to
remain plausible, the lack of strong absorption features appears to imply that either
super-Eddington outflows are not present in these sources, which if true would
strongly argue agaist the highly super-Eddington interpretation, or that if they are
present they must be directed away from our line of sight to the central accretion
flow. If weaker outflows are detected in the future with more sensitive observations
of these sources, this would suggest that the latter scenario is not the case.

\subsection{Emission Lines}

Until this point, we have been discussing the potential presence of the strong outflows
associated with highly super-Eddington accretion specifically assuming the ejected
material impinges our line of sight to the central accretion flow. We now consider the
possibility that strong super-Eddington outflows are present in Holmberg\,IX X-1 and
NGC\,1313 X-1 but, despite the large solid angles such outflows are expected to cover,
are directed away from our line of sight. In this scenario, one might expect to see
evidence of emission features as the outflowing material absorbs and reprocesses X-rays
from the central source. Narrow iron emission features, particularly from neutral iron,
are ubiquitously observed in the X-ray spectra of Galactic HMXBs, as the strong stellar
winds from the massive companions reprocess the X-rays from the central black hole (see
\citealt{Torrejon10}). As with HMXBs, the stellar companions of ULXs are also expected
to be fairly young, relatively massive stars (confirmed in the case of Holmberg\,IX X-1
by \citealt{Grise11}) which should naturally eject strong stellar winds regardless of
the nature of the compact object powering the ULX.

Again considering the immediate Fe K band, we find very similar constraints to those
placed on absorption features; any narrow emission features present must have equivalent
widths less than $\sim$30 and $\sim$50\,eV for Holmberg IX X-1 and NGC\,1313 X-1
respectively. Furthermore, emission features should not display the significant velocity
shifts possible for absorption features, so in principle we can be more precise about
the exact energies considered (note that any additional peculiar velocities of these
sources within their host galaxies could lead to the rest frame energy scale adopted
here being systematically offset, but such effects are unlikely to be large). Assuming
the redshifts quoted in Table \ref{tab_obs} are appropriate for these sources as well as
their host galaxies, the 99 per cent limits on emission features obtained for the
transitions highlighted in Figs. \ref{fig_hoix} and \ref{fig_ngc1313} (neutral,
helium-like and hydrogen-like iron) are $EW <$ 23, 25 and 12\,eV for Holmberg\,IX X-1,
and $EW <$ 41, 37 and 11\,eV for NGC\,1313 X-1 respectively. 

From these equivalent width limits, we can obtain upper limits for the reprocessing
column density using the relation derived by \cite{Kallman04}: $EW = AN_{\rm H}$, where
$A$ is some constant of proportionality. \cite{Kallman04} showed that for neutral iron,
$A_{\rm neutral} \simeq 3$ (for a solar iron abundance, equivalent widths in eV and
column densities in $10^{22}$\,\atpcm), which has been observationally confirmed for
Galactic HMXBs by \cite{Torrejon10}. Using this relation, (although correcting for the
$\sim$0.5\,\zsun\ metallicities of the host galaxies) we find that $N_{\rm H,neutral}
\lesssim$ 14 and 24 $\times 10^{22}$\,\atpcm\ for Holmberg\,IX X-1 and NGC\,1313 X-1
respectively, placing both at the lower end of the distribution observed in Galactic
HMXBs, which ranges from $\sim$10$^{22} - 10^{24}$\,\atpcm, with the majority
displaying \nh\ $\gtrsim 10^{23}$\,\atpcm; see Fig. 5 in \citealt{Torrejon10}. However,
that work also reported an anti-correlation between the neutral iron equivalent width
and HMXB luminosity, which they interpret as being due to an increase in ionisation of
the stellar winds in which the accreting black holes are embedded. The limits obtained
here for neutral iron are consistent with this picture. Any super-Eddington outfows
present can not provide any substantial neutral column in addition to the stellar wind
expected from the companion, but given the high ionisation of sub-Eddington outflows in
BHBs, we would naturally expect any super-Eddington outflows to be at least moderately
ionised.

If the outflows have a similar ionisation to the outflows observed in GRS\,1915+105
($\log\xi \sim 4$), iron will be primarily in its hydrogen-like ionisation state.
Re-calculating the constant of proportionality $A$ for Fe {\small XXVI}, taking the
threshold absorption cross-section of hydrogenic iron to be $\sim$9.5 $\times
10^{-21}$\,cm$^2$ (\citealt{Verner95}), the iron abundance to be $\sim$1.6 $\times
10^{-5}$ (\eg \citealt{tbabs}; $Z \sim 0.5$\,\zsun), and the flourescent yield to be
$\sim$0.5, we find that $N_{\rm H,\log\xi \simeq 4} \lesssim$ 9 and 8 $\times
10^{22}$\,\atpcm\ for Holmberg\,IX X-1 and NGC\,1313 X-1 respectively. Therefore,
there does not appear to be any substantial column of highly ionised reprocessing
material either.

However, we return to the issue that the expected evolution of the ionisation state of
the wind with increasing Eddington ratio is not well known, owing to the uncertain
evolution of the combination $nR^{2}$. As discussed previously, an increase in ionisation
will increase the column density inferred from the limit on the emission from Fe {\small
XXVI} as a larger fraction of the iron nuclei become fully ionised. However, in comparison
to the line-of-sight (absorption) case, it is more observationally plausible for the
ionisation to have decreased in this scenario. Indeed, as the Eddington ratio increases,
standard theory predicts that the launch radius of the wind should also increase, and it
is not unreasonable to imagine the outflow density might also increase; both of these
effects could serve to counteract the increase in luminosity and result in an overall
decrease in ionisation. If super-Eddington outflows are sufficiently less ionised than
sub-Eddington outflows, they could enter a regime in which resonant trapping and Auger
line destruction is efficient, which would result in only very weak emission lines (see
\eg \citealt{Ross96}).

A further caveat to our consideration of emission lines is that the relation between
line equivalent width and column density derived by \cite{Kallman04} assumes a spherical
distribution for the reprocessing medium. For a constant column, a non-spherical outflow
distribution (\ie covering a solid angle $\Omega < 4\pi$) would reduce the $EW$ expected
for the reprocessed emission lines. Although a spherical distribution is certainly
appropriate for the consideration of reprocessing by any stellar winds from the
companion, we now consider whether it is also a reasonable assumption for strong
super-Eddington outflows. It is widely expected that as the Eddington ratio increases,
the scale height of the wind should also increase (see \eg \citealt{King09},
\citealt{Dotan11}), so as the Eddington ratio increases, the assumption of a spherical
wind distribution becomes more and more appropriate. Indeed, \cite{Dotan11} find that
the winds launched by super-critical accretion become spherical beyond $L/L_{\rm E}
\simeq 5$, although this is somewhat dependent on the assumed model parameters, \eg the
disc viscosity and the opacity parameterisation. Nevertheless, although by virtue of the
lack of strong absorption features any strong outflows cannot be truly spherical in
these cases, the high Eddington ratios ($\sim$10) inferred for black hole masses of
10\,\msun\ might well result in something approaching a spherical outflow geometry, so
the estimated limits on the columns obtained might not be unreasonable. We stress that
while reducing the solid angle covered by the outflow is a legitimate way in which to
potentially increase the limits on the column density of the outflow, doing so would also
imply a reduction in the true Eddington ratio.

\subsubsection{Broader Lines}

\begin{figure*}
\begin{center}
\rotatebox{0}{
{\includegraphics[width=470pt]{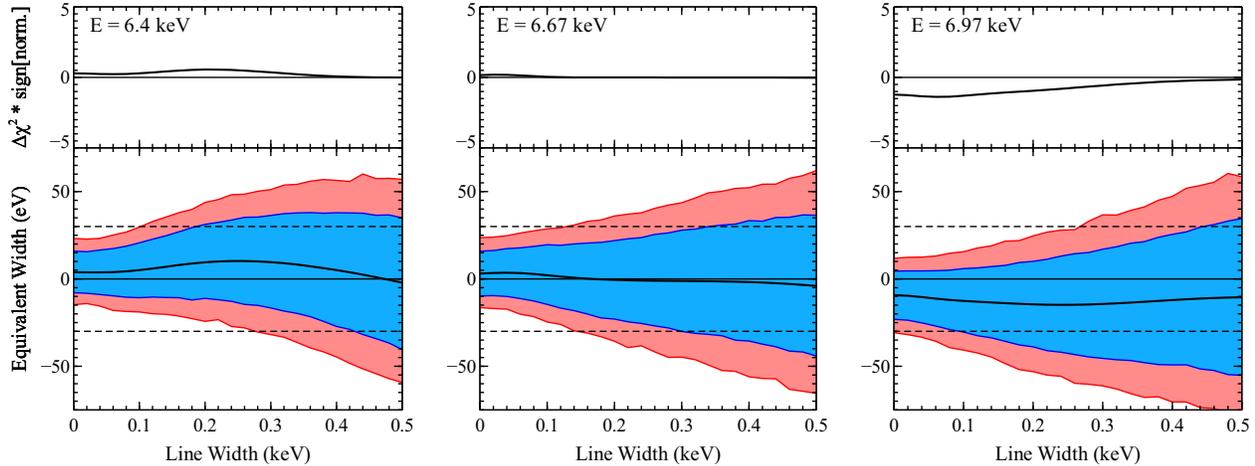}}
}
\end{center}
\caption{As in Fig. \ref{fig_hoix}, but here we plot the evolution in the $EW$ limits
with the assumed line width. For illustrative purposes, we only show evolution of the
limits obtained at the energies of the neutral, helium-like and hydrogen-like iron
transitions for Holmberg\,IX X-1. Larger equivalent widths are naturally allowed
as the line width increases, as by broadening the line the overall line flux can be
further increased without substantially perturbing the fit to the data.}
\label{fig_hoix_width}
\end{figure*}

In this work, we have focused primarily on narrow iron features, given the observed
absorption features from sub-Eddington outflows and the emission features from Galactic
HMXBs. However, the emission lines expected from an optically thick, diverging outflow
may be broadened by both Doppler shifts and electron scattering, particularly if the
outflow velocity is large (see \eg \citealt{Titarchuk09}, \citealt{Sim10}). Therefore,
we consider briefly the limits that can currently be placed on emission lines as a
function of the line width. For illustrative purposes, we demonstrate in Fig.
\ref{fig_hoix_width} how the $EW$ limits evolve over a range of assumed line widths for
the K\,$\alpha$ transitions of Fe {\small I, XXV} and {\small XXVI} for Holmberg\,IX
X-1. The limits naturally become weaker as the assumed width increases, therefore
stronger iron emission features could yet be present if they are substantially broadened
by the wind in which they arise. 

Unfortunately, the combination of the uncertainty in the ionisation state, solid angle
and the outflow velocity of the wind, all of which contribute towards determining the
width of any emission line produced by the outflow, means it is difficult to make any
clear predictions about the expected line widths, and hence provide a detailed
consideration of the limits on any such emission lines. We therefore conclude that at
this stage it is still too difficult to place strong constraints on the presence of
super-Eddington outflows that do not cover our line-of-sight just through consideration
of the expected iron emission.

\section{Conclusions}
\label{sec_conc}

Making use of all the good quality \xmm\ data publically available for two bright ULXs,
Holmberg\,IX X-1 and NGC\,1313 X-1, we have searched for (persistent) discrete atomic
features in their high energy spectra that could be associated with either iron emission
or absorption, and provide observational evidence for the massive outflows predicted if
these sources are accreting at substantially super-Eddington rates. We do not find any
statistically compelling evidence for any features, either in absorption or in emission.
However, we do find that the data currently available for these sources is of sufficient
quality that we can place fairly stringent limits on the strengths of any features
intrinsically present these spectra that remain undetected. Any features present
(absorption or emission) in the immediate Fe K energy band (6--7\,\kev) must have
equivalent widths weaker than $\sim$30\,eV for Holmberg\,IX X-1, and weaker than
$\sim$50\,eV for NGC\,1313 X-1.

In comparison to the strongest sub-Eddington outflows observed in GRS\,1915+105, which
imprint iron absorption features with equivalent widths of $\sim$30\,eV, the limits
obtained for the ULXs considered here appear quite restrictive, particularly when these
sources must be radiating at $\sim$5--10 times their Eddington limits if they host black
holes of similar masses, and should therefore be expelling at least 5--10 times as much
material as GRS\,1915+105. The difficulty in trying to reconcile these observational
constraints with the presence of strong line-of-sight outflows leads us to conclude that
either these sources do not launch such outflows, which would strongly argue against a
highly super-Eddington interpretation, or that they much be launched away from our
viewing angle. Additional deep observations of bright ULXs with current instrumentation,
and in particular with the micro-calorimeter due for launch on board \textit{Astro-H} in
the near future, will be essential for detecting any weak iron features in such sources,
and hence in placing stronger constraints on the nature of any outflows present, and the
ultimate nature of these sources.

\section*{ACKNOWLEDGEMENTS}

DJW acknowledges the financial support provided by STFC in the form of a PhD grant, and
ACF thanks the Royal Society. RCR thanks the Michigan Society of Fellows, and is supported
by NASA through the Einstein Fellowship Program, grant number PF1-120087. The figures
included in this work have been produced with the Veusz\footnote{http://home.gna.org/veusz/}
plotting package, written by Jeremy Sanders. This work is based on observations obtained
with \xmm, an ESA mission with instruments and contributions directly funded by ESA member
states and the USA (NASA). Finally, the authors would like to thank Tim Kallman for useful
discussions, as well as the referee for their feedback, which helped to improve the quality
and depth of this paper.

\bibliographystyle{mnras}

\bibliography{/home/dwalton/papers/references}

\appendix
\section{Individual Observation Details}
\label{appendix}

\begin{table*}
  \caption{Key parameters obtained for all the individual observations of Holmberg\,IX X-1
and NGC\,1313 X-1 with the phenomenological ultraluminous state continuum model (see text).}
\begin{center}
\begin{tabular}{c c c c c c c c c c c} 
\hline
\hline
\\[-0.3cm]
Source & Obs. & \nh & $kT_{\rm in}$ & $kT_{\rm e}$ & $\tau$ & \rchi\ ($\chi^{2}$/d.o.f.) \\
& & ($10^{21}$ cm$^{-2}$) & (keV) & (keV) & & &  \\
\\[-0.3cm]
\hline
\hline
\\[-0.25cm]
Holmberg\,IX\,X-1 & 0112521001 & $1.5\pm0.1$ & $0.24\pm0.04$ & $2.6^{+0.8}_{-0.4}$ & $7.5\pm1.2$ & 1.02 (3162/3102) \\
\\[-0.25cm]
& 0200980101 & & $0.22\pm0.05$ & $2.9^{+1.9}_{-0.5}$ & $6.6^{+1.0}_{-1.7}$ & \\
\\[-0.25cm]
& 0200980101 & & $0.24\pm0.01$ & $2.3\pm0.1$ & $9.2\pm0.4$ & \\
\\[-0.1cm]
NGC\,1313\,X-1 & 0106860101 & $2.6\pm0.1$ & $0.21\pm0.02$ & $2.2^{+0.4}_{-0.2}$ & $8.2\pm0.9$ & 1.05 (3968/3769) & \\
\\[-0.25cm]
& 0150280601 & & $0.30\pm0.06$ & 2* & $7.8^{+1.1}_{-0.7}$ & \\
\\[-0.25cm]
& 0150181101 & & $0.35^{+0.19}_{-0.15}$ & $1.6^{+1.2}_{-0.5}$ & $>5.3$ & \\
\\[-0.25cm]
& 0205230201 & & $0.25\pm0.05$ & 2* & $8.2^{+1.1}_{-0.8}$ & \\
\\[-0.25cm]
& 0205230301 & & $0.28^{+0.11}_{-0.06}$ & $1.6^{+0.2}_{-0.3}$ & $8.4^{+3.4}_{-1.2}$ & \\
\\[-0.25cm]
& 0205230601 & & $0.23^{+0.03}_{-0.02}$ & 2* & $9.0^{+0.6}_{-0.5}$ & \\
\\[-0.25cm]
& 0405090101 & & $0.22\pm0.01$ & $2.2^{+0.1}_{-0.2}$ & $8.5\pm0.5$ & \\
\\[-0.25cm]
\hline
\hline
\end{tabular}
\label{tab_comptt_all}
\end{center}
\end{table*}

\begin{figure*}
\begin{center}
\rotatebox{0}{
{\includegraphics[width=235pt]{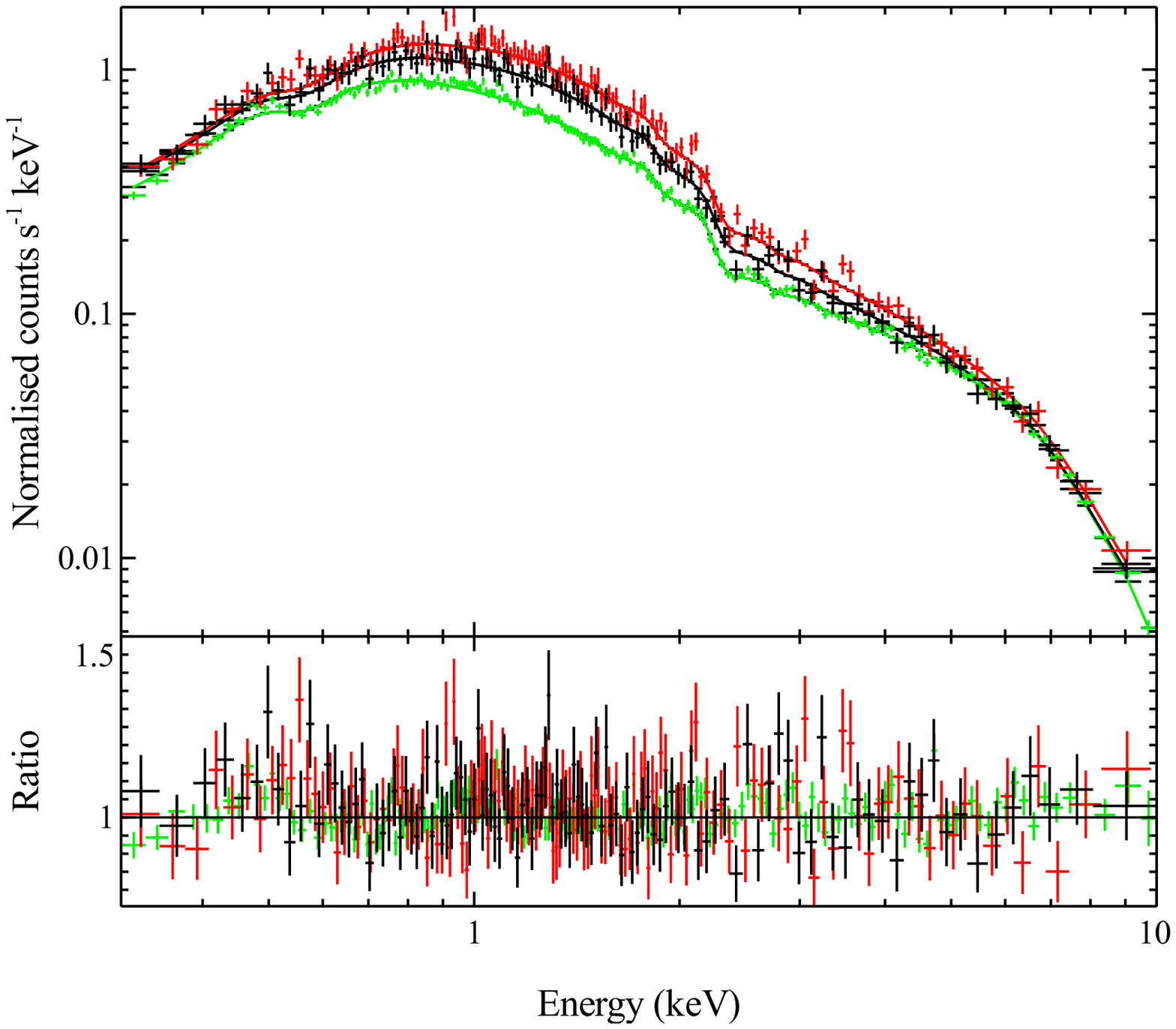}}
}
\hspace{0.25cm}
\rotatebox{0}{
{\includegraphics[width=235pt]{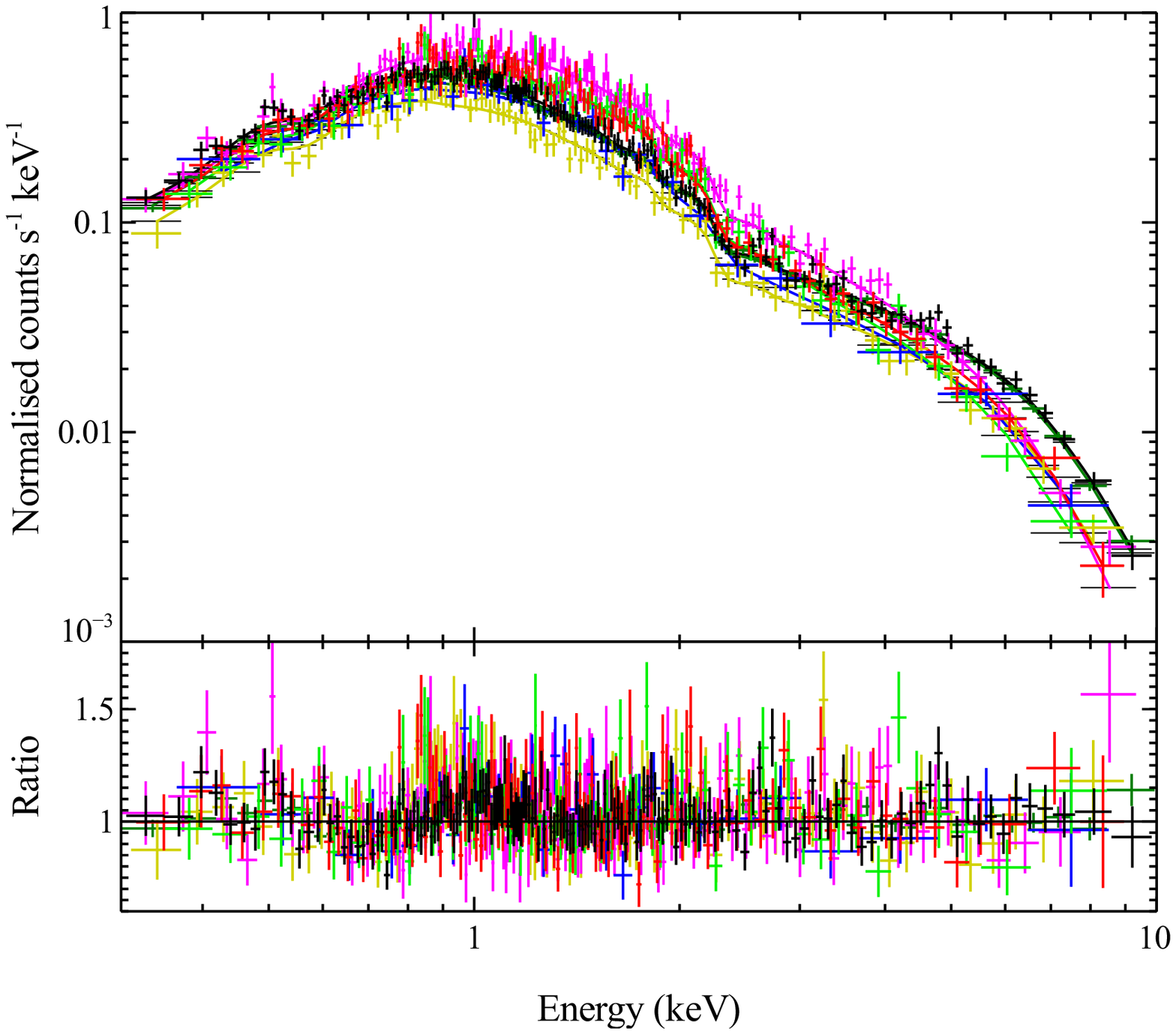}}
}
\end{center}
\caption{Count spectra for the individual observations of Holmberg\,IX X-1 (\textit{left
panel}) and NGC\,1313 X-1 (\textit{right panel}), and the data/model ratios for the best
fit continuum models presented in Table \ref{tab_comptt_all}. For clarity, only the
\epicpn\ data are shown, and the data are rebinned for display purposes only.}
\label{fig_comp}
\end{figure*}

Here, we examine briefly the individual observations of Holmberg\,IX X-1 and NGC\,1313 X-1
considered in this work to check whether it is reasonable to combine the data obtained into
a single spectrum in each case. For each source, we model the spectra obtained from the
individual observations simultaneously, adopting the same continuum model outlined in the
main body of the paper (see section \ref{sec_cont}). The key continuum parameters are all
free to vary between the observations, although we require that the neutral absorption
is the same for each. Owing to the highly varied data quality it was not always possible to
reliably constrain both the optical depth and electron temperature of the Comptonising
region in NGC\,1313 X-1, as these parameters often display some degeneracy. In these cases,
we fixed the electron temperature at 2\,\kev, similar to the values obtained for the other
observations. The results obtained are presented in Table \ref{tab_comptt_all}. Although
there are some minor differences in the parameter values obtained, they are all reasonably
similar for both sources.

In addition, to provide a visual comparison of the spectra from the various observations, we
plot the count spectra of the different observations for each source in Fig. \ref{fig_comp}.
This confirms the results obtained with the quantitative analysis presented in Table
\ref{tab_comptt_all}: there are some slight differences between the spectra of the
individual observations in terms of their high energy spectral slope/curvature, more so for
NGC\,1313 X-1 than Holmberg\,IX X-1, for which the high energy spectra are practically
identical, but ultimately the spectra are all fairly similar for both sources. Hence we
conclude that it is reasonable to combine them into a single, averaged spectrum for each
source in order to improve the high energy statistics and further facilitate our search for
narrow iron features.

As a final test, for each of the models obtained for the observations of NGC\,1313 X-1 (see
Table \ref{tab_comptt_all}) we inserted a narrow ($\sigma$ = 10\,eV) Gaussian absorption
line at 6.67\,\kev\ with an equivalent width of -50\,eV. Data were then simulated from each
model using the \epicpn\ response, keeping the respective exposures of the simulated spectra
the same as the good exposure times for the observations on which they are based (see Table
\ref{tab_obs}). We then combined the simulated datasets in the same manner as the real data,
and examined the characteristics of the absorption feature in the combined simulated dataset.
The narrow feature is well detected, providing an improvement of $\Delta\chi^{2} = 20$ for 2
additional degrees of freedom, and we find a line energy of $E = 6.65\pm0.04$\,\kev, and an
equivalent width of $EW = -52\pm19$\,eV (90 per cent confidence). Considering instead the
simulation of the long observation by itself, the absorption feature only offers an
improvement of $\Delta\chi^{2} = 14$ for 2 extra degrees of freedom. Therefore, even given
the mild variance in the high energy spectrum of NGC\,1313 X-1, combining the individual
spectra does aid the detection of persistent narrow features.

\label{lastpage}

\end{document}